\documentclass[10pt,letterpaper,twocolumn]{article} 

\usepackage{times,mathptmx}
\usepackage{ol2}
\usepackage{amsmath,amssymb}
\usepackage{color}

\begin{document}

\twocolumn[ 

\title{Third-order chromatic dispersion stabilizes Kerr frequency combs}

\author{Pedro Parra-Rivas$^{1,2}$, Dami\`a Gomila$^{2}$, Fran\c{c}ois Leo$^{3}$, St\'ephane Coen$^{4}$, and Lendert Gelens$^{1,5,*}$}

\affiliation{$^{1}$ Applied Physics Research Group, Vrije Universiteit Brussel, 1050 Brussels Belgium\\
$^{2}$ IFISC institute (CSIC-UIB), Campus Universitat de les Illes Balears, E-07122 Palma de Mallorca, Spain\\
$^{3}$ Photonics Research Group, Department of Information Technology, Ghent University, Ghent B-9000, Belgium\\
$^{4}$ Department of Physics,University of Auckland, Private Bag 92019, Auckland~1142, New Zealand\\
$^{5}$ Department of Chemical and Systems Biology, Stanford University School of Medicine, Stanford CA 94305, USA\\
$^{*}$ Corresponding author: lendert.gelens@vub.ac.be}

\date{\today}

\begin{abstract}%
  Using numerical simulations of an extended Lugiato-Lefever equation, we analyze the stability and nonlinear
  dynamics of Kerr frequency combs generated in microresonators and fiber resonators taking into account third-order
  dispersion effects. We show that cavity solitons underlying Kerr frequency combs, normally sensitive to
  oscillatory and chaotic instabilities, are stabilized in a wide range of parameter space by third-order
  dispersion. Moreover, we demonstrate how the snaking structure organizing compound states of multiple cavity
  solitons is qualitatively changed by third-order dispersion, promoting an increased stability of Kerr combs
  underlined by a single cavity soliton.
\end{abstract}

\ocis{230.5750, 190.5530, 190.3100, 190.4380}

] 


\noindent Optical frequency combs permit to measure light frequencies and time intervals with exquisite
accuracy, leading to numerous key applications. An octave of bandwidth is however typically required for
self-referencing \cite{udem_optical_2002, cundiff_colloquium_2003}. Kerr microresonators support on-chip generation
of such broadband frequency combs, with the potential to lead to very small footprints
\cite{kippenberg_microresonator-based_2011}. In fact, octave spanning ``Kerr combs'' have been demonstrated in both
silica microtoroids \cite{delhaye_octave_2011} and silicon nitride microresonators
\cite{okawachi_octave-spanning_2011}. With such large bandwidths, it is important to take third-order chromatic
dispersion (TOD) into account, which leads to asymmetric frequency combs. Comb generation in these conditions can be
modeled using a simple generalized mean-field Lugiato-Lefever equation (LLE), as recently shown in
\cite{coen_modeling_2013}. That study and others \cite{Chembo_LLE_2013} have highlighted a link between Kerr combs
and temporal cavity solitons (CSs) \cite{leo_temporal_2010}, but CSs are known to exhibit many
instabilities~\cite{leo_dynamics_2013}. Although the instabilities of Kerr frequency combs have been recently
intensely studied~\cite{matsko_chaotic_2013, Chembo_arxiv, Pedro_FC_2013, erkintalo_coherence_2014} and some effects
of fourth-order dispersion have been uncovered in the LLE \cite{Gelens_PRA_2008, GelensOL2010}, the influence of TOD
on the dynamics of CSs, and by association of Kerr combs, has not been investigated in detail. Addressing this
issue is the goal of the present Letter.

Here we show that TOD in microresonators can lead to suppression of dynamical regimes such as oscillations and
chaos, effectively stabilizing Kerr combs (a fact already hinted in \cite{milian_soliton_2014}). Furthermore, we
discuss the underlying dynamical mechanism behind this stabilization. We relate the dynamics of Kerr combs in the
presence of TOD to the snaking structure organizing single- and multi-peak CSs. Not only are single CSs stabilized
when introducing TOD, but multi-peak solutions are generally unstable and are only stable for large amounts of TOD.
Together this leads to an increased stability of Kerr combs with TOD, based on a stable underlying single CS.

\begin{table}[b]
  \vskip-4mm
  \centering
  \caption{Physical parameters and normalized TOD coefficient~$d_3$
           for three different optical systems. $\beta_2<0$ in all cases.}
  \begin{tabular}{l|ccc}
    \hline
                            & $\mathrm{MgF_2}$ \cite{Herr2014}
                            & $\mathrm{Si_3N_4}$ \cite{okawachi_octave-spanning_2011, coen_modeling_2013}
                            & Fiber \cite{Leo_prl_2013, jang_observation_2013}\\
    \hline
    $\alpha$                & $4.31\times 10^{-5}$ & $8.89\times 10^{-3}$ & $0.16$\\
    $L$                     & $6.2$~mm             & $628\ \mu$m          & 105~m\\
    $|\beta_2|$ (ps$^2$/km) & $5.9$                & $48.7$             & $0.1$--$0.54$\\
    $\beta_3$ (ps$^3$/km)   & $-0.35$             & $-0.14$            & $0.12$\\
    $d_3$                   & $-0.045$             & $-0.034$             & $0.18$--$2.2$\\
    \hline
  \end{tabular}
  \label{d3table}
\end{table}
The mean-field LLE that describes comb generation in microresonators has been described by many \cite{LL-1987,
matsko_mode-locked_2011, coen_modeling_2013, Chembo_LLE_2013, Chembo_arxiv, Pedro_FC_2013}. Using the normalization
of \cite{leo_temporal_2010}, and including dispersion up to third-order, that equation reads,
\begin{equation}
  \label{eq.1}
  \partial_{t}u=-(1+i\theta)u+i|u|^{2}u+u_{0}+i\partial_{\tau}^{2}u+d_3\partial_{\tau}^{3}u\,.
\end{equation}
Here $t$ is the slow-time describing the evolution of the intracavity field $u(t,\tau)$ at the scale of the cavity
photon lifetime, while $\tau$ is a fast-time that describes the temporal structure of that field along the resonator
roundtrip. The first term on the right-hand side describes cavity losses (the system is dissipative by nature);
$\theta$ measures the cavity detuning between the frequency of the input pump and the nearest cavity resonance; the
cubic term represents the Kerr nonlinearity, with the sign set so that it corresponds to the self-focusing case;
$u_0$ is the amplitude of the homogeneous (continuous-wave) driving field or pump; and the fast-time derivatives
model chromatic dispersion (dispersion is assumed anomalous at the pump frequency), with $d_3$ the relative strength
of the TOD. $d_3$ can be calculated from the physical parameters of the system, $d_3=(1/3) (2\alpha/L)^{1/2}
\beta_3/|\beta_2|^{3/2}$ \cite{xu_experimental_2014}, where $\alpha$ is half the percentage of power lost per
round-trip (the cavity finesse $\mathcal{F}=\pi/\alpha$), $L$ is the cavity length, and $\beta_2$ ($\beta_3$) is the
second (third) order dispersion coefficient. Table~\ref{d3table} shows typical values of these parameters and
corresponding relative TOD strength $d_3$ for three different physical systems, namely crystalline magnesium
fluoride (MgF$_2$) \cite{Herr2014} and silicon nitride (Si$_3$N$_4$) \cite{okawachi_octave-spanning_2011,
coen_modeling_2013} (see also \cite{Moss2013}) microresonators, as well as cavities made of a combination of
standard and dispersion-shifted optical fibers \cite{Leo_prl_2013, jang_observation_2013} and in which temporal CSs
in the presence of TOD have recently been observed \cite{jang_observation_2013}. The values span the whole parameter
range considered in this work.

\begin{figure}[t]
  \centering
  \includegraphics[scale=1.0]{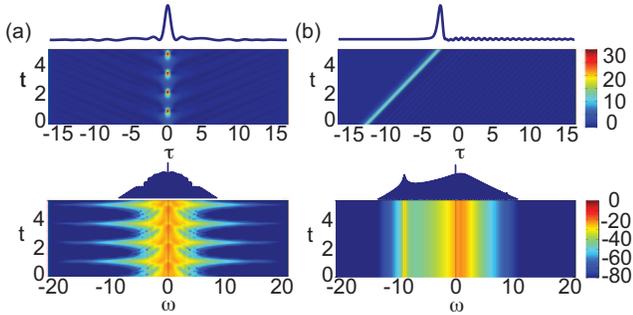}
  \vskip-1mm
  \caption{Evolution of (a) the temporal intensity profile of an oscillating CS over successive roundtrips (top) and
    its associated comb spectrum in dB (bottom) in the absence of TOD ($d_3=0$). (b) With $d_3=0.15$, the system is stable.
    The profiles at time $t=5$ are shown on top of each graph. $\theta=6.1$, $u_0=4$.}
  \label{oscillations}
  \vskip-2mm
\end{figure}
\begin{figure}[t]
  \includegraphics[scale=1.0]{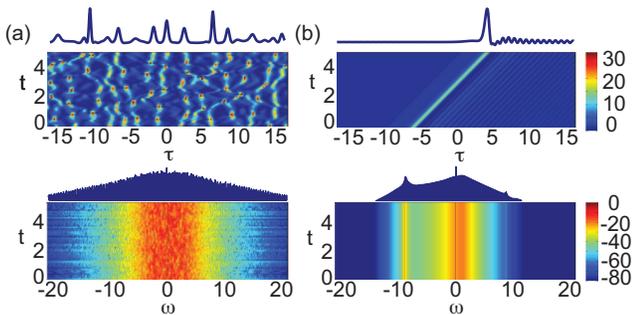}
  \vskip-1mm
  \caption{Same as Fig.~1 but for $u_0=5.5$, for which the solution exhibits spatial-temporal chaos in the absence of
    TOD.}
  \label{chaos}
  \vskip-5mm
\end{figure}
An example of oscillatory and chaotic behaviors of an isolated CS in the absence of TOD (second-order dispersion
only, $d_3=0$) is shown in Figs~\ref{oscillations}(a) and~\ref{chaos}(a), respectively. Only the pump amplitude
differs in these two simulations as indicated in the captions. The pseudocolor plots at the top show the evolution
of the temporal intensity profile of the intracavity field while the bottom ones are the corresponding spectra.
Figures~\ref{oscillations}(b) and~\ref{chaos}(b) reveal that when the magnitude of TOD is sufficiently large the
dynamical instabilities are completely suppressed: the CS is stable, albeit in a moving reference frame. Note that
this motion has no practical effect on the comb spectrum, which is perfectly steady. The TOD breaks the reflection
reversibility $\tau\rightarrow-\tau$, which leads to asymmetries in the temporal and spectral
profiles~\cite{mussot_optical_2008, Leo_prl_2013, jang_observation_2013, milian_soliton_2014}. This asymmetry is
also responsible from the observed constant-velocity temporal drift: the CS carrier frequency is shifted from the
pump due to spectral recoil from the emission of dispersive waves \cite{coen_modeling_2013, jang_observation_2013,
milian_soliton_2014}, which are clearly seen both in Figs~\ref{oscillations}(b) and~\ref{chaos}(b). Accordingly, the
group-velocity of the CS differs slightly from that of the pump. In fiber cavities, a similar change in
group-velocity occurs through acoustic effects and leads to long-range CS interactions \cite{jang_ultraweak_2013}.

\begin{figure}
\centering
  \includegraphics[scale=1.0]{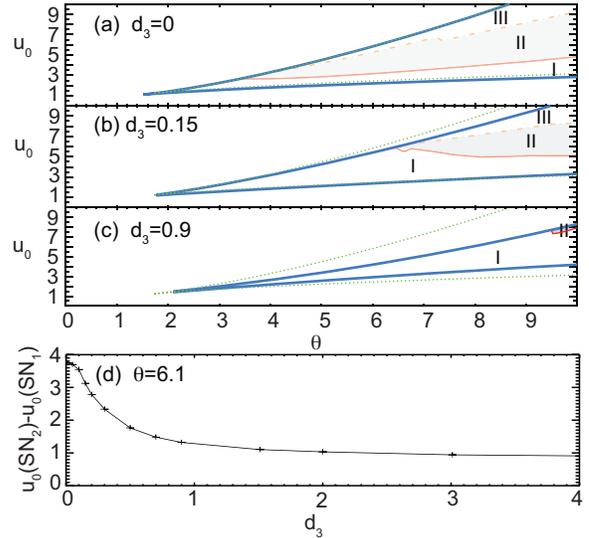}
  \caption{Regions of existence of CSs and their instabilities in the ($\theta$,$u_0$) parameter space for (a) $d_3 = 0$, (b)
    $d_3 = 0.15$, (c) $d_3 = 0.9$. Region I: stable CSs; region II (light-gray): time-oscillation solutions arising
    through Hopf bifurcation H (red line); region III: spatio-temporal chaos. The green dotted (blue solid) lines
    denote the SN bifurcations of the HSS (CSs), respectively. (d) Width of the pinning region where CSs
    exist versus $d_3$ for $\theta=6.1$.}
  \label{stability_u0_det}
\end{figure}
In order to verify whether the stabilization of the CS and corresponding comb is a general feature in the presence
of TOD, we analyzed the stability of CSs in the whole parameter space ($\theta$,$u_0$) for various values of the
TOD. The result of this analysis is shown in Fig.~\ref{stability_u0_det}. To interpret this figure, let us first
recall that the homogeneous steady state (HSS) $u_\mathrm{s}$ of Eq.~(\ref{eq.1}) is given by $I_\mathrm{s}
[1+(\theta-I_\mathrm{s})^2] = I_0$, where $I_\mathrm{s}=|u_\mathrm{s}|^2$ and $I_0=|u_0|^2$. For $\theta<\sqrt{3}$,
only one HSS exists, hence the system is monostable. For $\theta>\sqrt{3}$ three HSS states appear, one of which is
a saddle point (unstable), hence this regime is referred as bistable. These homogeneous solutions are connected
through saddle-node (SN) bifurcations shown as green dotted lines in Fig.~\ref{stability_u0_det}. On this figure we
have also indicated the cusp point~C at $\theta = \sqrt{3}$. On top of HSSs, it is well known that the LLE admits
extended periodic pattern solutions that appear through modulational instability (MI) \cite{LL-1987,
tlidi_control_2007}. When the detuning is sufficiently large ($\theta>41/30$ in the absence of TOD), these patterns
emerge subcritically from the HSS, hence the patterned solution and a stable HSS coexist. It is this generalized
bistability that makes possible the existence of multiple stationary temporal CSs in the LLE. They coexist with the
patterned solutions within a so-called snaking or pinning region delimited by two SN bifurcation points
\cite{Champneys, Coullet00}. These points, referred to as $\mathrm{SN}_1$ and $\mathrm{SN}_2$, are plotted using
thick blue solid lines in Figs~\ref{stability_u0_det}(a)--(c) for increasing values of TOD. In the absence of TOD, a
detailed overview of the regions of multistability between the HSSs, CSs, compound CS states, and extended patterns
can be found in \cite{leo_dynamics_2013, Chembo_arxiv, Pedro_FC_2013}. In these works, the organization of various
dynamical regimes, including oscillations and chaos, was also discussed.

From Fig.~\ref{stability_u0_det}, it is clear that while the region of existence of the HSSs is independent of TOD,
the snaking region in which CSs can be found (between the blue lines) shrinks with increasing values of the TOD. To
highlight this point, we plot in Fig.~\ref{stability_u0_det}(d) the width of the snaking region
[$u_0(\mathrm{SN}_2)-u_0(\mathrm{SN}_1)$] versus the TOD strength $d_3$ for a fixed detuning $\theta=6.1$. Here it
can be seen that the shrinkage, while initially rapid, somewhat saturates at higher $d_3$ such that a region
admitting CS solutions can be found independent of the TOD strength $d_3$.

Figures~\ref{stability_u0_det}(a)--(c) also illustrate the dependence of various regions of instabilities of a
single CS as a function of the TOD strength $d_3$. CSs are stable in region~I while they are found to exhibit a
time-oscillatory behavior [as was illustrated in Fig.~\ref{oscillations}(a)] in region~II (light-gray colored).
These oscillatory solutions emerge through a Hopf bifurcation~H (thin red line). In the absence of TOD, this Hopf
bifurcation has theoretically been demonstrated to originate in a Gavrilov-Guckenheimer codimension-2 point
\cite{Pedro_FC_2013} and has been experimentally observed using fiber resonators \cite{leo_dynamics_2013}. Above
region~II, for increasing values of pump power and detuning, we find that the temporal evolution of the CSs lead to
spatio-temporal chaos [as was illustrated in Fig.~\ref{chaos}(a)]. The part of parameter space where this chaotic
behavior is located is referred to as region~III. Figs~\ref{stability_u0_det}(a)--(c) demonstrate that both the
oscillatory (II) and chaotic (III) regions of instabilities shrink and shift to higher values of the detuning
$\theta$, confirming that the stabilization of CSs and Kerr combs in the presence of TOD, which was exemplified in
Figs~\ref{oscillations} and \ref{chaos}, is a general feature.

\begin{figure}[t]
  \centering
  \includegraphics[scale=1]{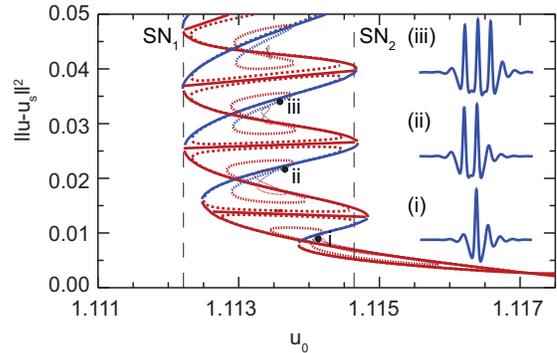}
  \vskip-5mm
  \caption{Snaking bifurcation diagram showing the energy of single- and multi-peak CS solutions for $\theta=1.5$ versus
    pump amplitude $u_0$. Plots are made for increasing values of TOD, $d_3=0$ (solid line), and $d_3= 0.01$, $0.05$, $0.075$, with increasingly shorter and lighter dashed curves. Blue (red) curves are stable (unstable). Insets (i)--(iii), related to the corresponding points on the snaking diagram, are
    examples of temporal intensity profiles of CSs obtained for $d_3=0.05$.}
  \label{snaking_lowdetuning}
  \vskip-5mm
\end{figure}

The dynamical regimes discussed above only concern Kerr combs underlined by a single CS. However, in the absence of
TOD, multistability between many different stationary solutions is known to exist \cite{Pedro_FC_2013}. These
solutions consist of multiple CSs and can be understood as bound states of single CSs. We therefore proceed to
studying the effect of TOD on the stability and bifurcation structure of multi-peak solutions.
Fig.~\ref{snaking_lowdetuning} shows the typical bifurcation structure of CSs plotted in terms of their energy, for
$d_3=0$ (solid lines) and $d_3>0$ (dashed lines). Energy is calculated with the homogeneous background $u_\mathrm{s}$
removed, i.e., using the norm $||u-u_\mathrm{s}||^2=\int |u(\tau)-u_\mathrm{s}|^2 d\tau$. Blue (red) lines are
stable (unstable) solutions, respectively. Detuning is fixed at $\theta=1.5$. Without TOD, $d_3=0$, two branches of
CSs bifurcate from the HSS together with the extended pattern at the MI point (zero norm point at the bottom right
of the Figure). These two branches are related to solutions with, respectively, an odd and even number of peaks.
Initially these solutions correspond to small amplitude unstable states. Following these branches ``snaking''
upwards (increasing the norm), these states grow in amplitude and energy, and successively gain and lose stability
through SN bifurcations. At the SN points, extra peaks are also added symmetrically on both sides of the temporal
structures resulting in new bound states of CSs of higher norm (see insets in Fig.~\ref{snaking_lowdetuning}). The
even and odd state branches are connected through a branch corresponding to different kind of structures called rung
states \cite{Burke09}. The back and forward oscillation of the branches is referred to as homoclinic snaking or a
snakes-and-ladders structure \cite{Knobloch1}. This snaking or pinning region is defined by the asymptotic location
of the SNs higher up the snaking structure ($\mathrm{SN}_{1,2}$). The dependence of the pinning region on the
various system parameters was shown in Fig.~\ref{stability_u0_det}. However, as revealed by
Fig.~\ref{snaking_lowdetuning}, TOD not only changes the size of the pinning region, but also alters the whole
snaking structure of CSs.

As the $\tau$-reversibility symmetry is broken by TOD, the bifurcations responsible for the rung states become
imperfect and the snaking structure breaks up. This occurs in two different fashions depending on the detuning
$\theta$. For low values of the detuning, as exemplified in Fig.~\ref{snaking_lowdetuning}, the break-up leads to a
stack of isolas. Examples of temporal intensity profiles associated to such isolas for $d_3=0.05$ are shown in
inset. The creation of isolas is well known in conservative systems harboring localized structures when breaking
reversibility \cite{Burke09}. For a fixed value of the detuning, these isolas shrink with increasing values of
$d_3$, until they eventually disappear when the cusp of CSs (the point C where $\mathrm{SN}_1$ and $\mathrm{SN}_2$
meet in Fig.~\ref{stability_u0_det}) moves beyond the actual value of the detuning. At this point the system no
longer admits CS solutions.
\begin{figure}
\centering
\includegraphics[scale=1]{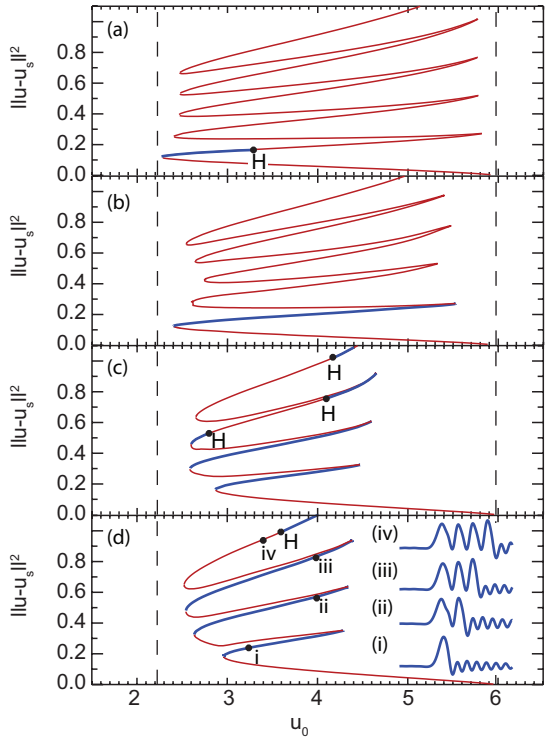}
 \caption{
Snaking diagrams of the single and multi-peak CS branches for $\theta=6.1$ for increasing values of TOD, $d_3 =
0.1$, $0.15$, $0.6$, and $0.9$ [from (a) to (d)]. Solid (dashed) lines correspond to stable (unstable) solutions. H:
Hopf bifurcations.
}
 \label{snaking_highdetuning}
\end{figure}

For higher values of the detuning $\theta$, which are more relevant to practical Kerr comb generation, the situation
is different. This case is illustrated in Fig.~\ref{snaking_highdetuning} that shows the snaking structure for a
detuning $\theta=6.1$ and increasing strength of the TOD. In this regime all the multi-peak CSs organized in the snaking structure are unstable in the absence of TOD and for small values of TOD. CS branches corresponding to even and odd numbers of peaks merge in a type of ``mixed snaking.'' A similar transition between isolas and mixed snaking has recently been studied in detail in
the context of the Swift-Hohenberg equation \cite{Sandstede_2013}. Figs~\ref{snaking_highdetuning}(b)--(d) show that
TOD increasingly stabilizes the multiple peak solutions, starting with the one peak branch (b), and then gradually
stabilizing the two-peak one (c), three-peak one (d), etc.  This stabilization process seems to involve multiple
Hopf bifurcations as most clearly seen in Fig.~\ref{snaking_highdetuning}(c). Examples of typical solutions are
shown in inset of Fig.~\ref{snaking_highdetuning}(d), showing an increased amplitude of the oscillatory tails. For
TOD values corresponding to microresonators (see Table~\ref{d3table}), this stabilization process may however not be
sufficiently strong. The detuning is typically ramped up in experiments (i.e., going from
Fig.~\ref{snaking_lowdetuning} to Fig.~\ref{snaking_highdetuning}), and in that process the single-peak CS solution
may typically become the preferred remaining stable solution. The results presented here therefore could help explain the
emergence of a single CS observed both experimentally and theoretically in recent works exploring the route to
stable Kerr frequency combs \cite{Herr2014, lamont_route_2013}. We finally remark that other solutions (not shown here) such as multiple displaced single CSs connected via their oscillatory tails can also exist and we aim to investigate their bifurcation structure and stability in future work.

In summary we have shown that the stability, dynamics, and bifurcation structure of CSs and Kerr combs is largely
modified in the presence of TOD. TOD tends to suppress dynamical instabilities of the single CS such as oscillations
and chaos. Moreover the so-called snaking structure, organizing the single and multiple CS solutions, is altered by
TOD. Our analysis has revealed that despite multi-peak solutions can be stabilized by TOD, such stabilization
requires an increasing amount of TOD as the number of peaks increases. Altogether, by promoting their increased
stability, these TOD effects thus especially lead to Kerr combs underlined by a single CS.

This research was supported by the Research Foundation - Flanders (L.G. and P. P.-R.), by the Belgian American Educational Foundation (L.G.), by the Spanish MINECO and FEDER under Grant INTENSE@COSYP (FIS2012-30634), by Comunitat Autonoma de les Illes Balears, by the Belgian Science Policy Office (BelSPO) under Grant No. IAP 7-35, and by the Marsden Fund of the Royal Society of New Zealand (S.C.).

\clearpage

\section*{References with title}

\end{document}